\documentclass{emulateapj}
\usepackage{epsfig}

\newcommand{\thalf}{t_{1\!/2}}
\newcommand{\tE}{t_{\rm E}}

\newcommand{\qmax}{Q_{\rm max}}
\newcommand{\Dfm}{\Delta F_{\rm max}}

\begin{document}
\title{Microlensing Surveys of M31 in the Wide Field Imaging Era}
\author{Edward~A.~Baltz}

\affil{KIPAC, Stanford University,
P.O. Box 20450, MS 29, Stanford, CA 94309, {\tt eabaltz@slac.stanford.edu}}

\begin{abstract}
The Andromeda Galaxy (M31) is the closest large galaxy to the Milky Way, thus
it is an important laboratory for studying massive dark objects in galactic
halos (MACHOs) by gravitational microlensing.  Such studies strongly complement
the studies of the Milky Way halo using the the Large and Small Magellanic
Clouds.  We consider the possibilities for microlensing surveys of M31 using
the next generation of wide field imaging telescopes with fields of view in the
square degree range.  We consider proposals for such imagers both on the ground
and in space.  For concreteness, we specialize to the SNAP proposal for a space
telescope and the LSST proposal for a ground based telescope.  We find that a
modest space-based survey of 50 visits of one hour each is considerably better
than current ground based surveys covering 5 years.  Crucially, systematic
effects can be considerably better controlled with a space telescope because of
both the infrared sensitivity and the angular resolution.  To be competitive, 8
meter class wide-field ground based imagers must take exposures of several
hundred seconds with several day cadence.

\end{abstract}

\keywords{gravitational lensing --- galaxies: halos ---
galaxies: individual (M31) --- dark matter}

\section{Introduction}

Almost twenty years ago Paczy\'nski (1986) proposed that dark objects in the
Milky Way halo might be found by searching for the rare gravitational
microlensing of stars in nearby galaxies: the Large and Small Magellanic Clouds
(LMC and SMC) and also M31.  This technique could thus be used to explore the
dark matter problem in searching for these objects, dubbed MACHOs.  Both
primordial black holes (a candidate for non-baryonic dark matter, known to
dominate the clustering matter in the universe) and stellar remnants (a
candidate for baryonic dark matter, an unknown component accounting for roughly
half of the baryons today) are interesting MACHO candidates.  For a dark halo
consisting of solar--mass MACHOs, of order one in one million LMC stars is
being lensed (with a magnification of 30\% or more) at any given time.  These
events are transient, developing over typically 100 days.  A monitoring
campaign could hope to detect these events, thus shedding light on the nature
and composition of the Milky Way halo.  Several groups have performed such
campaigns over more than ten years, and have extended the technique to more
distant galaxies as well.

The MACHO collaboration (Alcock et al.~2000) monitored the LMC for more than 8
years, concluding from the 5.7 year dataset that there is a population of
lenses making a contribution of approximately 20\% in objects of roughly
half-solar mass.  The EROS collaboration (Afonso et al.~2003) has monitored the
LMC and SMC over a similar time period, and finds only an upper limit of 25\%
on the microlensing component.  The MACHO collaboration result is intriguing in
that while the possibility that the dark matter consists solely of these
lensing objects is ruled out, there may be a subdominant non-baryonic dark
matter component.  Otherwise, the result speaks to the dark baryon problem.

More than a decade ago, Crotts (1992) pointed out that the Andromeda Galaxy
(M31) would be an excellent target for a microlensing survey, in particular
because of the asymmetry expected in the rate of halo lens events due to the
steep inclination of the M31 disk.  Both the Milky Way and M31 halos could be
studied in detail.  Subsequently, this result was discussed in depth (Jetzer
1994; Baltz \& Silk 2000; Kerins et al.\ 2001; Baltz, Gyuk \& Crotts 2003).

Microlensing surveys of M31 require image subtraction, as very few stars can be
resolved from the ground.  This is the regime of ``pixel'' lensing (Crotts
1992; Baillon et al.~1993; Tomaney \& Crotts 1996; Gould 1996).  Several
collaborations, including MEGA (de Jong et al.~2004) (preceded by the
VATT/Columbia survey (Crotts \& Tomaney 1996, Uglesich 2001, Uglesich et
al.~2004)), AGAPE (Ansari et al.~1999, Auri\`ere et al.~2001, Calchi Novati et
al.~2002, Paulin-Henriksson et al.~2003, Calchi Novati et al.~2003), and WeCAPP
(Riffeser et al.~2003), have produced a number of microlensing event candidates
involving stars in M31.  The results of the VATT/Columbia survey (Uglesich et
al.~2004) indicate the presence of a lensing halo at the 20\% level, excluding
the case of only self lensing at greater than 95\% confidence.

Gould (1995) noted that the Hubble Space Telescope (HST) could perform a
microlensing survey of the Virgo cluster using M87.  Such a survey has been
carried out (Baltz et al.\ 2004), placing only an upper limit that the Virgo
halo is less than 60\% lenses in the solar mass range.

The current observational situation is uncertain.  Ground based programs are
moving forward, such as SuperMACHO, MEGA, and AGAPE.  To understand populations
of dark objects in galactic halos, wide field imaging telescopes currently
being proposed and studied would have large advantages.  In this paper we
discuss the potential of wide field imaging for microlensing surveys of M31.
For significant progress to be made, ground-based surveys would need to be
performed with 8 meter class telescopes, or with night-long integrations on
smaller telescopes.  We discuss using the proposed LSST for this purpose.  A
space-based wide field survey offers many advantages relative to the ground; we
discuss using the proposed SNAP space telescope for microlensing.  Graff \& Kim
(2001) have previously discussed using the SNAP telescope for a microlensing
survey of the LMC.

\section{Microlensing Rate}
\label{sec:theory}
Point gravitational lenses always exhibit two images of any source.  The
separation of these images for solar mass lenses within the Local Group tends
to be microarcseconds, thus the term microlensing.  Even with unobservable
image splitting, the magnification can be large.  For lenses and sources in
relative motion, the magnification is transient, resulting in a microlensing
event.  In this section we describe the calculation of the rate of detectable
microlensing events.

\subsection{Basics}

The lens equation can be written in terms of the Einstein radius and angle,
which for a point mass lens are given by
\begin{equation}
R_{\rm E}=\sqrt{\frac{4GM}{c^2}\frac{D_{\rm l}D_{\rm ls}}{D_{\rm s}}},\;\;\;
\theta_{\rm E}=\sqrt{\frac{4GM}{c^2}\frac{D_{\rm ls}}{D_{\rm s}D_{\rm l}}},
\end{equation}
where $M$ is the lens mass, $D_{\rm l}$ is the distance to the lens, $D_{\rm
s}$ is the distance to the source, and $D_{\rm ls}$ is the distance between the
lens and source.  The lens equation relating source position $\theta_S$ and
image position $\theta_I$ is then given by (in angular coordinates with the
lens at the origin)
\begin{equation}
\vec{\theta}_S=\vec{\theta}_I\left(1-\frac{\theta_{\rm
E}^2}{\theta_I^2}\right).
\end{equation}
This equation can be written with all angles written in units of the Einstein
angle, in particular $u=\theta_S/\theta_{\rm E}$.  The magnification was first
given by Einstein (1936):
\begin{equation}
A(u)=1+f(u^2),\;\;\;f(x)=\frac{2+x}{\sqrt{x(4+x)}}-1.
\end{equation}
For $u\ll1$, $A\approx 1/u$, thus large magnifications are possible for point
sources.

A microlensing event has a characteristic timescale called the Einstein time,
$\tE=R_{\rm E}/v_\perp$, given in terms of the perpendicular velocity of the
lens relative to the source $v_\perp$.  For rectilinear motion, we find
\begin{equation}
u^2(t)=\beta^2+\frac{(t-t_0)^2}{\tE^2},
\end{equation}
where $\beta$ is the minimum impact parameter of the lens, in units of the
Einstein angle.  For a star with unlensed flux $F_\star$, the microlensing
lightcurve is then
\begin{equation}
F(t)=F_\star+F_\star f(u^2(t))=F_\star+F_\star\delta(u(t)).
\end{equation}
The Einstein time can be easily extracted from the lightcurve if $F_\star$ is
known.  The lightcurve shape contains some information on $\beta$ (and thus
$F_\star$), though the dependence is weak, requiring very high quality data
(e.g.\ Paulin-Henriksson et al.~2003).

In the previous discussion we have assumed point sources.  If the angular size
of the source $(\theta_\star=R_\star/D_{\rm l})$ is large, namely
$\theta_\star>\beta\theta_{\rm E}$, finite source size effects can be
significant, as summarized by Yoo et al.~(2004).

\subsection{``Pixel'' Microlensing}
The unlensed flux of the source star is difficult to measure.  Significant
blending is possible even for nearby sources (e.g.\ in the LMC).  Most sources
in M31 are highly blended, even with the resolution available from a space
telescope.  The full width at half maximum timescale, rather than the Einstein
timescale, is the one usually measured.  It is given by
$\thalf=2\,\tE\,w(\beta)$, where
\begin{equation}
w(\beta)=\sqrt{2f(f(\beta^2))-\beta^2},
\end{equation}
with limiting behavior
\begin{equation}
w(\beta\ll1)=\beta\sqrt{3},\;\;\;w(\beta\gg1)=\beta\sqrt{\sqrt{2}-1}.
\end{equation}
We find that $\thalf\sim \tE\,\beta$ for all values of $\beta$.

The degenerate form (high magnification limit $\beta\rightarrow0$) of the
microlensing lightcurve is
\begin{equation}
F(t)=B+\Dfm\left[1+12\left(\frac{t-t_0}{\thalf}\right)^2\right]^{-1/2},
\label{eq:lc1}
\end{equation}
where $B$ is the baseline flux and $\Dfm=F_\star\delta(\beta)\approx
F_\star/\beta$ is a fit parameter expressing the maximum {\em increase} in flux
from the lensed star.  In the absence of blending $B=F_\star$.  The ``pixel''
lensing regime in where sources are only resolved when they are lensed.  The
measured parameters are $\Dfm$ and $\thalf$.  Without knowing $F_\star$,
measuring the Einstein time $\tE$ requires measuring $\beta$, which provides
only a small correction to the degenerate lightcurve.  High quality data is
required (Paulin-Henriksson et al.~2003).  We proceed using the distribution of
microlensing events in $\thalf$, though not as useful as that in $\tE$.  We
note that $\Dfm\thalf\sim F_\star \tE$ in the high magnification regime, so
Einstein times can be recovered statistically, given that the distribution of
$F_\star$ is known (Gondolo 1999).  This distribution in $F_\star$ is simply
the stellar luminosity function, and at least the first non-trivial moment is
known: the surface brightness fluctuation (SBF) magnitude.

In practice the lightcurves are extracted by image differencing.  A reference
image is constructed from some number of individual exposures, and is then
subtracted from the individual images.  In this way the baseline is removed,
and the only remaining sources are variable.

\subsection{Microlensing Rate}
The timescale distribution for microlensing events is expressed as the integral
along the the line of sight of the lens density times the cross section (Griest
1991; Baltz \& Silk 2000):
\begin{equation}
\frac{d\Gamma_0}{d\tE}=\frac{4D_sv_c^2}{M_{\rm lens}}\int_0^1dx\, \rho_{\rm
lens}\,\omega_{\rm E}^4e^{-\omega_{\rm E}^2-\eta^2}I_0(2\omega_{\rm E}\eta),
\label{eq:simplerate}
\end{equation}
where $D_l=xD_s$, $\omega_{\rm E}=R_{\rm E}/(v_c\tE)$, $I_0$ is a modified
Bessel function of the first kind, $v_c$ is the circular velocity of the lens
population $(v_c=\sigma\sqrt{2})$, and $\eta=v_t/v_c$ is the transverse
velocity of the line of sight relative to the lens population.  Lenses are
assumed to be drawn from a maxwellian velocity distribution.  The transverse
velocity due to the motion of source and observer, assuming that the source
velocity is prescribed and not taken from a distribution (e.g.\ disk sources
whose random velocities are much less than their bulk velocities in the halo
frame).  This equation does not account for detection efficiency, which can be
expressed as a probability to detect an event with certain fit parameters
$P[\thalf,\qmax,\beta]$, where $\qmax\propto\Dfm$ is just the maximum flux
expressed as a signal-to-noise ratio.  The rate of detectable events is then
just
\begin{equation}
\frac{d\Gamma}{d\tE}=\int_0^\infty d\beta\,\frac{d\Gamma_0}{d\tE}\,
P\left[2\tE w(\beta),Q_\star\delta(\beta),\beta\right].
\label{eq:prate}
\end{equation}
Note that the fit parameters $\thalf$ and $\Dfm$ have been replaced with more
physical ones depending on $\beta$.  Here, $Q_\star$ ($\propto F_\star$) is the
naive (photon counting) significance with which the source star is detected if
blending is ignored.  Since $d\Gamma_0/d\tE$ is independent of $\beta$, the
rate can be written as
\begin{equation}
\frac{d\Gamma}{d\tE}=\beta_{\rm eff}
\left(\tE,Q_\star\right)\,\frac{d\Gamma_0}{d\tE}\left(M_{\rm lens}\right).
\label{eq:rate}
\end{equation}
The distribution is integrated over $\tE$, and then over the mass function of
lenses and the luminosity function of sources to arrive at the total observable
rate.  Note that $\tE$ may be unknown event by event, but integrating over all
$\tE$ yields the total observed event rate.  The distribution $d\Gamma/d\thalf$
can be derived similarly,
\begin{equation}
\frac{d\Gamma}{d\thalf}=\int_0^\infty d\beta\,\frac{d\Gamma_0}{d\tE}
\left(\frac{\thalf}{2w(\beta)}\right)\,
\frac{P\left[\thalf,Q_\star\delta(\beta),\beta\right]}{2w(\beta)}.
\end{equation}

We include finite source effects using a simple prescription (Baltz \& Silk
2000).  For a given source, we can determine the required magnification to give
a lensed flux large enough to pass a detection criterion.  For a given source
diameter $\theta_\star$, the maximum magnification is a function of $x$: as
$x\rightarrow 1$, $A_{\rm max}\rightarrow 0$.  We can solve for the largest $x$
allowing the required magnification, and truncate the $x$ integral in
equation~\ref{eq:simplerate} accordingly.  A quick estimate shows that at most
a handful of events might be observed toward M31 with significant finite source
effects.

We have assumed that the source velocities have no random component.  If the
source velocities are maxwellian (as should be approximately the case for a
bulge), the velocity integrals can be separated out.  The outcome is the same
if the identification $v_c^2({\rm lens})\rightarrow v_c^2({\rm
lens})+x^2v_c^2({\rm source})$ is made.

\subsection{Monte Carlo Event Rates}

For a given microlensing survey, we can perform a Monte Carlo simulation to
determine the event detection probability $P[\thalf,\qmax,\beta]$.  For this
work it will be safe to assume that the Poisson errors in photon counting will
be approximately gaussian, i.e.\ the number of photons expected is
significantly larger than 1.

The time sampling in a given survey is used to generate simulated microlensing
lightcurves with fit parameters $\thalf$, $\qmax$, $\beta$.  Random peak times
$t_0$ are chosen.  Data points with Poisson errors are generated for each time
sample.  The generated lightcurves are then passed through a set of criteria
for detection.  In this way, the probability $P$ is calculated for a set of fit
parameters.  This process is repeated, generating a table of probabilities as a
function of the fit parameters, to be used in equation~\ref{eq:prate}.

In this work we use the following detection threshold.  A total of 10$\sigma$
is required, collected over $N$ consecutive samples that are
$10\sigma/\sqrt{N}$ above the baseline.  We let $N=2$, 3, 4, 5, 8, allowing for
short bright events as well as longer dimmer events.

\section{Modeling of M31}
\label{sec:modeling}

We use the microlensing model of M31 described in Baltz, Gyuk \& Crotts
(2003).  This model consists of sources in the bulge and disk of M31, and
lenses in the bulge, disk, and halo of M31, as well as Milky Way halo lenses.

\subsection{Galactic Halos}

We model the dark halos of the Milky Way and M31 as cored isothermal spheres,
having asymptotic $1/r^2$ density profiles, giving flat rotation curves.
Lenses making up a fraction $f_{\rm MW,M31}$ of the dark halo have a space
density given by
\begin{equation}
\rho(x,y,z) = f\,\frac{v_c^2}{4 \pi G}
\frac{\sqrt{1-q^2}/\left(q\,\sin^{-1}\sqrt{1-q^2}\right)}{x^2+y^2 + (z/q)^2
+r_c^2},
\end{equation}
where $v_c$ is the asymptotic rotation speed, $r_c$ is the core radius, and $q$
is the flattening, with $q=1$ indicating no flattening.  We assume that
$v_c=220$ km s$^{-1}$ for the Milky Way halo and $v_c=240$ km s$^{-1}$ for the
M31 halo.  We vary the core radius for M31 as $r_c=$0.1, 1, 2, 5, 10 kpc, and
take flattening values $q=1$ and $q=0.3$.  We assume that the halo velocity
distribution is maxwellian.  We impose halo cutoffs at a distance of 200 kpc
from the respective centers.

\subsection{Stellar Populations}

Sources in M31 are taken to reside in a two component model consisting of a
double-exponential disk (Walterbos \& Kennicutt 1988; Gould 1994) and a bulge
(Kent 1989).  Details of this model have been discussed previously (Baltz, Gyuk
\& Crotts 2003).  Self lensing between these components is computed (all cases:
bulge-bulge, bulge-disk, disk-bulge, and disk-disk), serving as the background
for the detection of a lensing halo.  The M31 disk and bulge are then
considered as sources for halo lensing events.

\subsection{Mass and Luminosity Functions}

We adopt the Chabrier (2001) mass function for stellar lenses in M31, both in
the disk and the bulge.  For halo lenses we assume a delta function mass
function, though varied between 10$^{-3}M_\odot$ and 10 $M_\odot$.  For the
luminosity function of stars, we consider the bulge and disk separately.  For
the disk, we use the $R$-band (for ground based surveys) and $J$-band (for
space based surveys) luminosity functions of Mamon \& Soneira (1982).  The
bulge is more complicated, as data are scarce.  We take an average (in $\log
dN/dM$) of $V$-band and $I$-band data for $M_R>0$ (using Terndrup, Frogel \&
Whitford~(1990) for $I$-band and Holtzman et al.~(1998) for $V$-band, both
toward the Milky Way bulge).  For $M_R<0$ we attach a power law slope of 0.59
taken from the MACHO project data (Alves 2001).  For the $J$-band, we use an
isochrone for an old metal-poor population (Lauer \& Mighell 2003) with surface
brightness fluctuation magnitude $\overline{M}_J=-3.43$, which is adequate for
this study.

The full dataset of a microlensing survey would allow for better measurements
of the stellar luminosity functions, at least in terms of the SBF magnitude.  A
space based survey would obviously be considerably better for this purpose than
a ground based one, simply due to the considerably better resolution.

\section{Sample Wide Field Surveys}

From the microlensing model of M31, we now proceed to study wide field surveys.
Some experimental parameters are needed, namely the sensitivity and angular
resolution of the imaging systems in question.  In the following, we discuss a
ground based plan and a space based plan.

\subsection{Ground Based Survey with LSST}

The Large Synoptic Survey Telescope (LSST) is a proposed instrument for mapping
the entire accessible sky every several days (Tyson et al.\ 2004).  As such, it
will discover microlensing in M31 and elsewhere.  We assume some basic
experimental parameters as follows.  The LSST focal plane is very large, of
order 10 square degrees.  All of the interesting region of M31 is easily
covered in one field.  One caveat is that the site of LSST is as yet undecided,
and the visibility of M31 from the southern hemisphere is limited.  However,
the discussion carries over to any wide-field imaging system with exposure
times scaled appropriately.

The LSST aperture is an annulus, with an effective diameter of 6.5 meters.  One
possible plan for the LSST survey is to scan the entire visible sky in 4 colors
every 4 days, with 10 second exposures, possibly in pairs.  We will assume an
$R$-band photometric zero point of $m_R^{\rm ZP}=27.4$ mag (the flux giving one
photoelectron per second), implying a 10 second exposure zero point of
$m_R^0=29.9$ mag, giving one photoelectron over the exposure.  Furthermore, we
assume that the nose level is twice the photon counting noise, as is typically
achievable in ground based microlensing of M31 (the degradation is possibly due
to unresolved variable stars).  We use a cadence of 4 days to do the Monte
Carlo simulation described previously.  This is an oversimplification, as bad
weather is not taken into account, but acceptable for our purposes.

The resolution can be characterized by one number, usually taken to be the
seeing.  More useful for microlensing is $\Omega_{\rm PSF} = 1/\sum \psi_i^2$,
where $\psi_i$ is the normalized PSF (Gould 1996).  For a gaussian PSF with
seeing (full width at half maximum) $d_0$, it is easy to show that $\Omega_{\rm
PSF}=\pi d_0^2/\ln4$.  We assume that the typical seeing at the LSST site
(still being discussed) will be 0.75\arcsec, giving $\Omega_{\rm PSF}=1.27$
square arcseconds.

We take the surface brightness $\mu_R$ from the M31 model (Baltz, Gyuk \&
Crotts 2003).  The background sky brightness is assumed to be uniform, with
$\mu_{\rm sky}=21$ mag arcsec$^{-2}$ in the $R$-band.  Lastly, we assume that
the distance modulus to M31 is $D=24.5$, corresponding to a distance of 795
kpc.  We can now express the conversion to signal-to-noise ratio as follows,
for a star with apparent magnitude $m_R$ magnified by a factor $A=1+\delta$,
\begin{equation}
Q=\frac{1}{2}\frac{10^{-0.4\left(m_R-m_R^0/2\right)}\;\delta}
{\sqrt{\Omega_{\rm PSF}\left(10^{-0.4\mu_R}+10^{-0.4\mu_{\rm sky}}\right)}},
\end{equation}
and the factor of $1/2$ is due to the fact that the noise level is twice the
photon counting noise, usually the case for observing M31 from the ground.
This is likely due to unresolved variable stars, e.g.\ RR Lyrae (Crotts 2004).
This expression assumes that the noise in the reference image used in image
subtraction is much less than the noise in individual images, as is usually the
case.

\subsection{Space Based Survey with SNAP}

The SuperNova Acceleration Probe (SNAP) is a proposed wide field space
telescope chiefly aimed at measuring the acceleration of the universe by Type
Ia supernovae and by weak gravitational lensing (Aldering et al.\ 2004).

The SNAP aperture is planned to be 2 meters in diameter.  The focal plane
covers 0.7 square degrees, though in a strange shape.  There are 9 colors, with
filters mounted directly on the focal plane.  There are 4 optical and 4
infrared sections.  The optical sections are broken up into a 6x6 grid, with 6
filters, while the infrared sections are broken up into a 3x3 grid, with 3
filters.  The current plan is for the 9 filters to start at $B$-band (central
wavelength 4400 \AA), redshifting by a factor of 1.15 going redward, the
longest filter having a central wavelength of 1.346 $\mu$m.  This range covers
the usual color system $BVRIJ$ and slightly redder.  Microlensing rates tend to
be largest in the reddest filters, thus we will assume that $J$-band is used
for the primary event detection.  Converting the quoted sensitivity (Aldering
et al.\ 2004) from AB magnitudes, we use $m_J^{\rm ZP}=23.7$ mag, and for a 100
second exposure (the lower limit based on bandwidth) $m_J^0=28.7$ mag.  We
conservatively estimate that $\Omega_{\rm PSF}=0.1$ square arcseconds.

A microlensing survey with SNAP would take time away from the main surveys, but
the resources required are modest.  To cover the central region of M31, 27
pointings would be adequate.  This would cover each position in the 3 infrared
filters and 3 of the optical filters.  The optical coverage could be staggered,
so that in two visits, all points would be covered in the 3 infrared filters
twice and the 6 optical filters once.  Bandwidth to the telescope limits the
exposure times to be more than about 100 seconds, so we assume 100 second
exposures.  This means a visit would consist of 2700 seconds of integration,
plus overheads which may be as large as 50\%.  M31 is at low ecliptic latitude
(+33.3$^\circ$), thus even from space it is only visible for part of the
year.\footnote{As pointed out by Graff \& Kim (2001), the LMC is an attractive
target for a microlensing survey with SNAP for just this reason -- it lies
within 5$^\circ$ of the south ecliptic pole.}  Constraints from the roll angle
of the telescope would likely limit the visibility to 2 periods of 90 days in a
year, when the sun is 90$^\circ$ away from M31.  We consider surveys of 90
days, with cadences of 2 days and 4 days (in the optical filters 4 days and 8
days).

We adapt the M31 model to $J$-band simply estimating that $R-J=1.67$ for the
disk and $R-J=1.72$ for the bulge.  We assume a sky brightness $\mu_J=21.7$ mag
arcsec$^{-2}$.  Because many stars in M31 are resolved at this resolution, we
must be careful in estimating the noise level.  We take a conservative
approach, adding the star brightness to the background.  A more correct
approach would divide the surface brightness into resolved and unresolved
components, thus there would be no double counting.  We connect to
signal-to-noise ratio similarly to the ground-based case,
\begin{equation}
Q=\frac{10^{-0.4\left(m_J-m_J^0/2\right)}\;\delta}
{\sqrt{10^{-0.4m_J}+\Omega_{\rm PSF}
\left(10^{-0.4\mu_J}+10^{-0.4\mu_{\rm sky}}\right)}}.
\end{equation}

The resolution of a space-based telescope such as SNAP provides an important
advantage to a microlensing survey of M31 in that some of the stars will be
resolved.  For such stars, the Einstein timescale can be measured directly.  We
use a conservative criterion for determining which stars are resolved: if a
star is 10 times brighter than the surface brightness contained in $\Omega_{\rm
PSF}$, then it is resolved.  This can be expressed as
\begin{equation}
m<\mu-2.5-2.5\log\Omega_{\rm PSF}\rightarrow m<\mu,
\label{eq:sb_te}
\end{equation}
where $m$ is the apparent magnitude of the source, and $\mu$ is the surface
brightness of the galaxy (in the same band).

\subsection{Number of Events in Mock Surveys}

For the sample surveys in both LSST and SNAP we provide the expected number of
events.  These are listed in Table~\ref{tab:events}.  We compute rates for
flattenings $q=1$ and $q=0.3$, and for halo core radii $r_c=$0.1, 1, 2, 5, 10
kpc.  We use a delta function for the halo lens mass function, varying between
$10^{-3}M_\odot$ and $10 M_\odot$, in intervals of 1/8 dex.
Table~\ref{tab:events} illustrates the fiducial case $q=1$, $r_c=2$ kpc,
$M=-1/2$ dex solar.  We also compute the distribution in timescales for these
events, both $d\Gamma/d\tE$ and $d\Gamma/d\thalf$.  In the case of SNAP, we
separately calculate the number of events with $\tE$ determination according to
equation~\ref{eq:sb_te} and those with only $\thalf$ determined.

\begin{deluxetable}{lcccc}
\tablewidth{0pt}
\tablecaption{Expected Number of Events\label{tab:events}}

\tablecomments{Expected number of events in mock surveys.  We assume a core
radius for the halos $r_c=2$ kpc and no flattening ($q=1$).  The MACHO mass is
fixed at $-1/2$ dex solar (.32 $M_\odot$).  Numbers for the individual halos
assume a 100\% MACHO fraction.  The SNAP survey covers 90 days, while the LSST
numbers are given per full year of observations -- for M31 this would likely
take 3 years.  In addition to the baseline 10 second exposure for an LSST
survey, we also give the rates for identical surveys (4 day cadence) taking 100
second and 1000 second exposures as well.  The LSST numbers can easily be
scaled to different telescopes, e.g. an unobscured 3.2 meter telescope has 1/4
the area, thus the numbers are appropriate for 40 second, 400 second, and 4000
second exposures.}

\tablehead{survey & self & MW & M31 & self + 20\% halos}

\startdata
SNAP 2 day & 165 & 240 & 635 & 340\\
SNAP 4 day & 110 & 185 & 470 & 240\\
LSST 10 s & 18 & 33 & 79 & 40\\
LSST 100 s & 99 & 155 & 400 & 210\\
LSST 1000 s & 465 & 590 & 1590 & \hspace{1ex}900
\enddata
\end{deluxetable}

\section{Maximum Likelihood}

Having calculated the expected number of microlensing events in various
surveys, we would like to understand how well the parameters of the
microlensing model can be measured.  We use a maximum likelihood technique
first developed for the VATT/Columbia survey (Uglesich et al. 2004), as we
describe below.

\subsection{Simulated Datasets}

\begin{figure*}
\epsfig{file=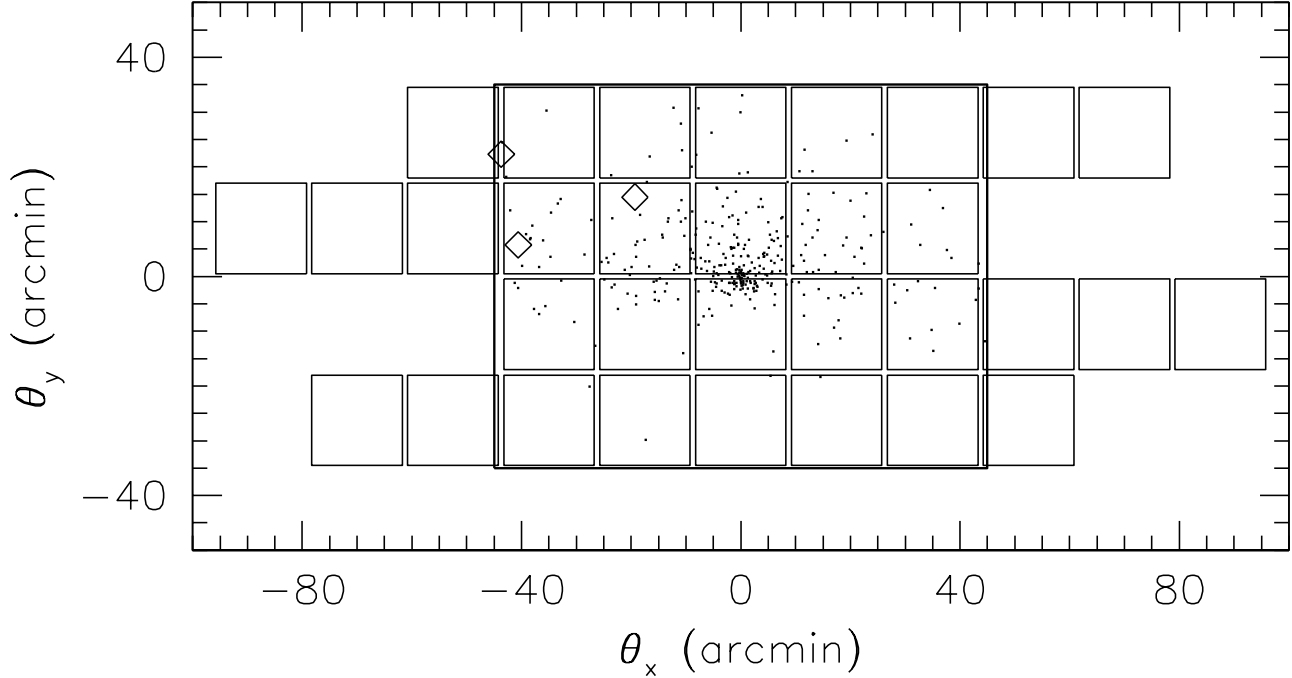, width=\textwidth}
\caption{SNAP focal plane with simulated dataset, assuming a halo lens fraction
of 20\% with mass $-1/2$ dex solar.  The cadence is 2 days, with a total survey
duration of 90 days.  The positions of the infrared detectors are indicated,
showing the scan across the field.  The infrared detectors are individually 1/3
the width of the illustrated squares -- each square thus requires 3 exposures
in the scan.  With the edge effects, 26 exposures are required to cover the
central 4x5 chip region, which is boxed.  To stagger the position in
alternating scans for the optical filters (whose positions are the mirror image
of those shown), one additional exposure is required, making 27.  Dots mark the
simulated events, and diamonds mark those few simulated events passing the
$\tE$ measurement criterion.  Note the considerable asymmetry between the near
(bottom) and far (top) sides.}
\label{fig:focalplane}
\end{figure*}

\begin{figure*}
\epsfig{file=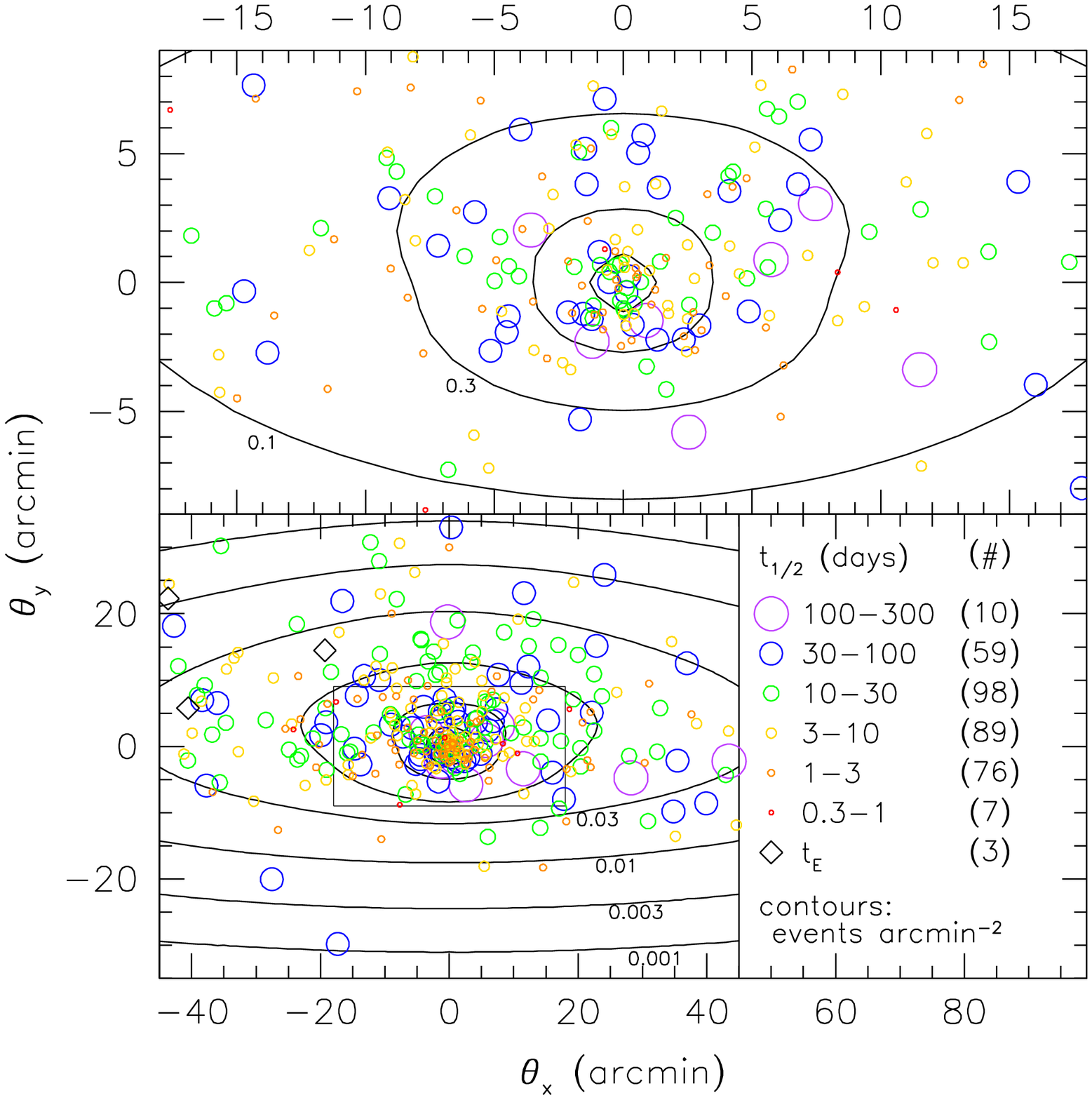, width=\textwidth}
\caption{Detail of simulated dataset shown in Figure~\ref{fig:focalplane},
along with contours of expected event density.  Again, the halo lens fraction
is 20\% and the lens mass is $-1/2$ dex solar.  The bottom panel shows the full
simulated region, while the top panel shows the M31 bulge region.  Circles
indicate simulated events, with timescales in days of (smallest to largest,
with colors online only) 0.3d $<\thalf<1$d (red), 1d $<\thalf<3$d (orange), 3d
$<\thalf<10$d (yellow), 10d $<\thalf<30$d (green), 30d $<\thalf<100$d (blue),
100d $<\thalf<300$d (purple).  Diamonds indicate events where $\tE$ could be
measured.  The contours indicate expected events per square arcminute, and are
spaced by 1/2 decade (so ``3'' means $\sqrt{10}$ in the contour labels).
Again, the near--far asymmetry is clear from the contours.}
\label{fig:colorfocalplane}
\end{figure*}

We simulate surveys of M31 based on the expected event rate as calculated in
the previous section.  The event rates are binned in position and log timescale
with intervals of 1 arcminute and 1/8 decade respectively, for each component
of the total lensing rate.  For a given set of parameters, such as lens mass
and halo lens fraction, a total expected lensing rate $\tau_i$ is constructed
in each bin across the field of view and in timescale.  Simple Poisson
statistics are then used to generate a number of events in each bin (usually
zero as there are many more bins than expected events).  The simulated data is
then a set of number of events in bins, $\{n_i\}$.  It is simple to generate
many realizations for each set of parameters.  In fact we generate two parallel
sets of bins, for the expected event rate with and without a measurement of
$\tE$.  In Figure~\ref{fig:focalplane} we illustrate one such realization, for
a halo lens fraction of 20\% and lens mass $-1/2$ dex solar $(0.316 M_\odot)$ .
We assume the 2 day cadence SNAP survey of 90 day duration, and show the size
of SNAP focal plane in the infrared filters (the optical filters are the mirror
image).  Note that we have separated source stars into two groups according to
equation~\ref{eq:sb_te}, and then separately generated realizations for events
with measured $\tE$ and those with only $\thalf$ measured.  In
Figure~\ref{fig:colorfocalplane} we show this simulated dataset in more detail,
both illustrating the timescales of events, and also focusing on the central
region, which has the highest event density, and where star-star lensing
dominates.

\subsection{Likelihood Function}

The likelihood function is clearly just the product of a Poisson distribution
for each bin, for a number of events $n_i$, where the expectation is $\tau_i$,
\begin{equation}
{\cal L}=\prod_i e^{-\tau_i}\frac{\tau_i^{n_i}}{n_i!}.
\end{equation}
The expected number of events in each bin $\tau_i$ is a function of the
parameters of the lensing model.  The most important are lens mass and halo
lens fraction.  The likelihood function is maximized over these parameters.  We
can define $\chi^2=-2\,\ln \cal L$, though we don't expect this quantity to be
distributed in the usual way as this regime of Poisson statistics is far from
gaussian,
\begin{equation}
\chi^2=2\sum_i\left[\tau_i-n_i\ln\tau_i+\ln (n_i!)\right].
\end{equation}

For each set of input parameters (lens mass, lens fraction, core radius,
flattening), we compute the likelihood function $\cal L$.  Marginalizing over
halo shape (core radius and flattening), we construct contours of equal
likelihood.  We integrate the likelihood function over the lens mass and halo
lens fraction to identify the contours containing probabilities corresponding
to the usual $1\sigma$ (68.269\%), $2\sigma$ (95.450\%), $3\sigma$ (99.730\%),
and $4\sigma$ (99.994\%) confidence levels.  We also marginalize over each axis
in turn to arrive at the one-dimensional confidence intervals for lens mass and
halo lens fraction.

\begin{figure*}
\epsfig{file=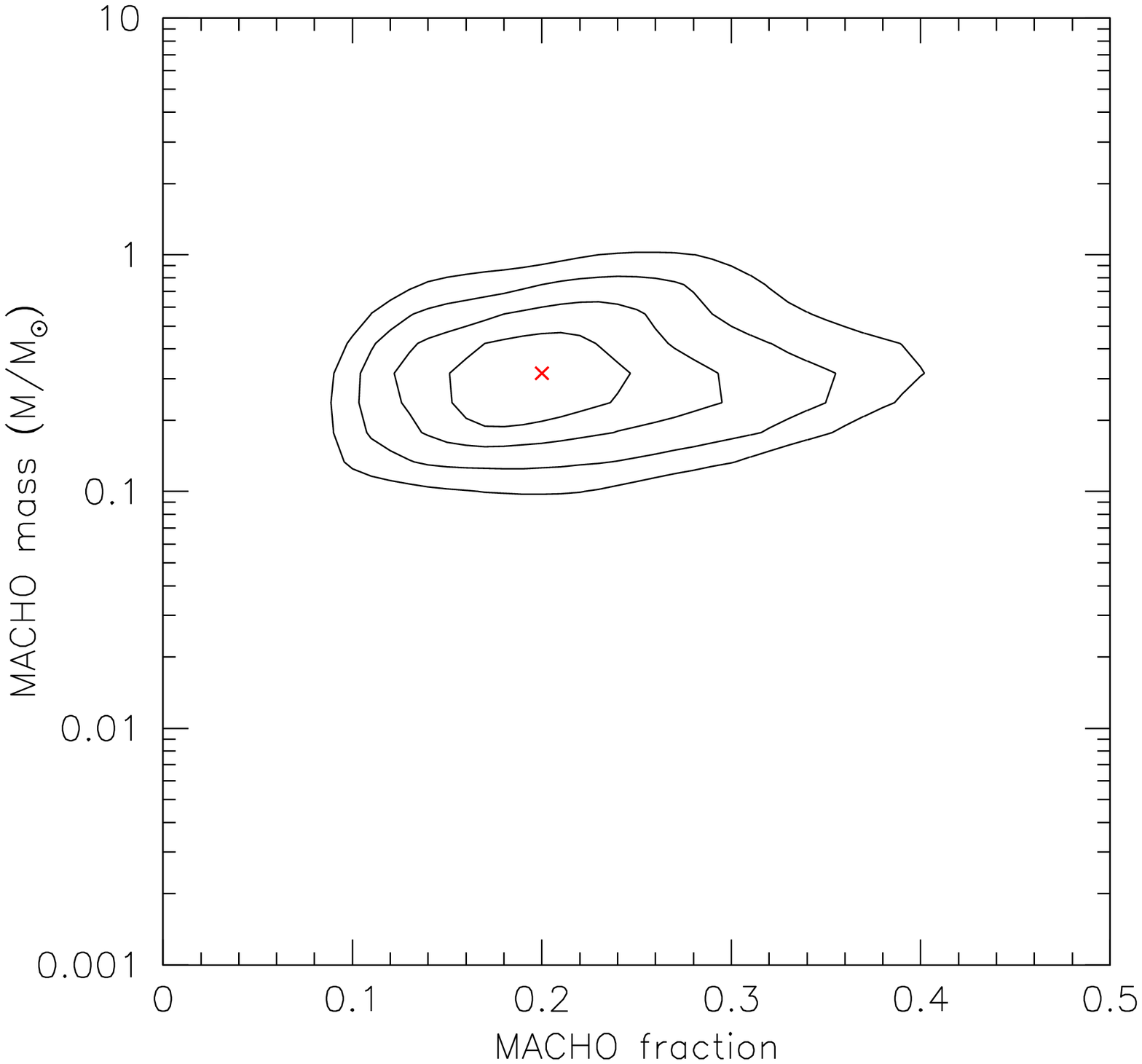, width=0.5\textwidth}
\epsfig{file=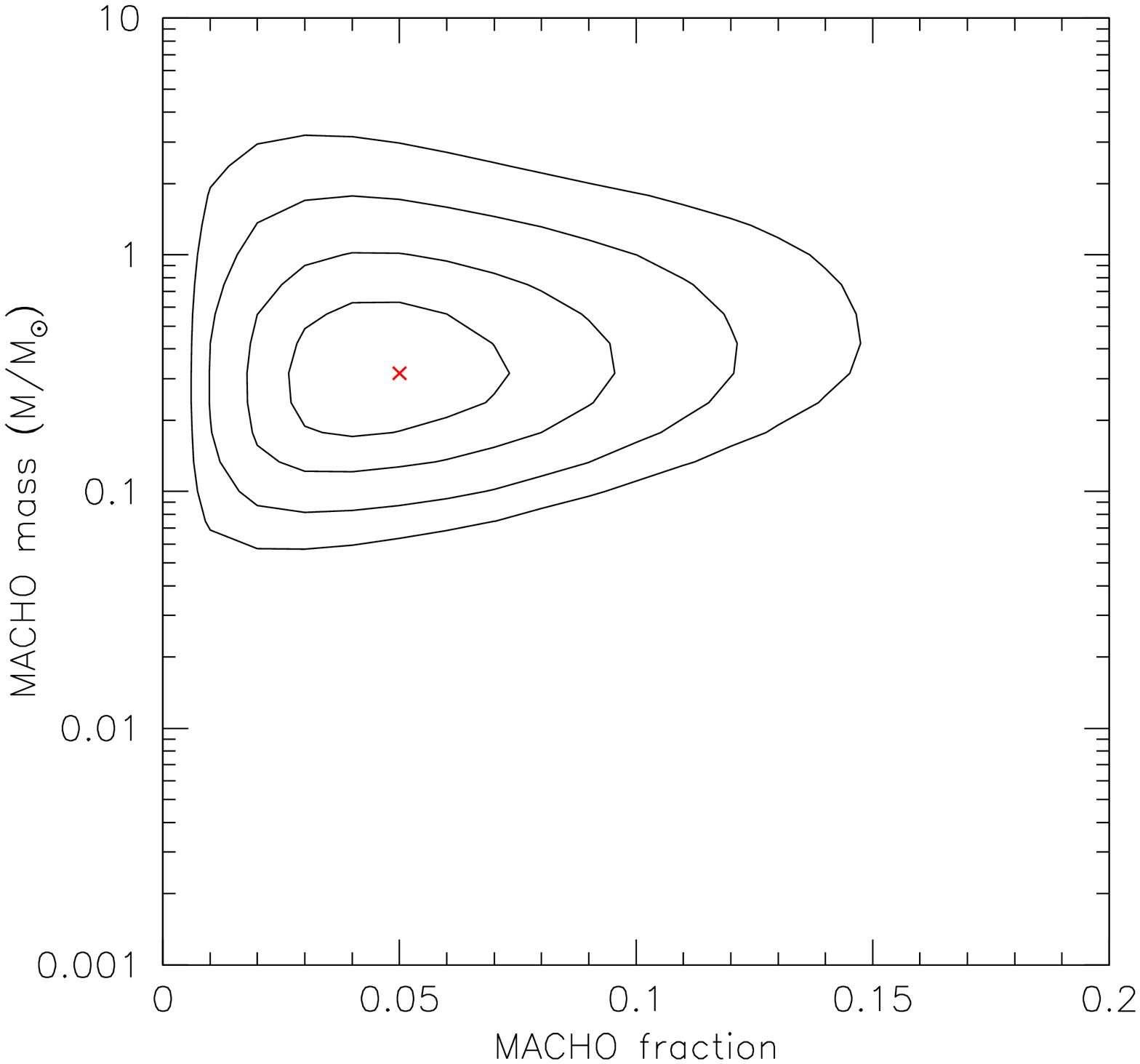, width=0.5\textwidth}
\epsfig{file=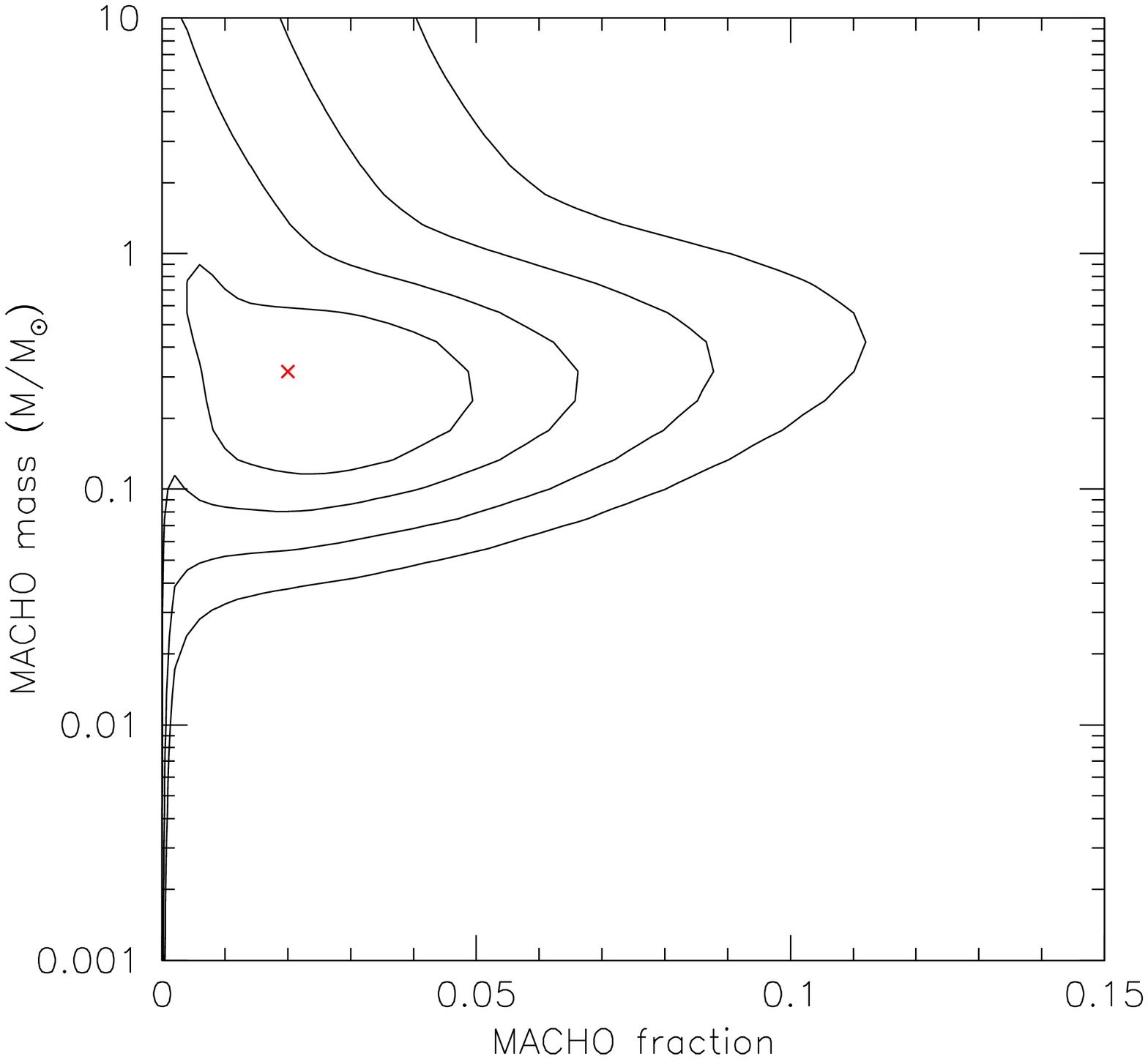, width=0.5\textwidth}
\epsfig{file=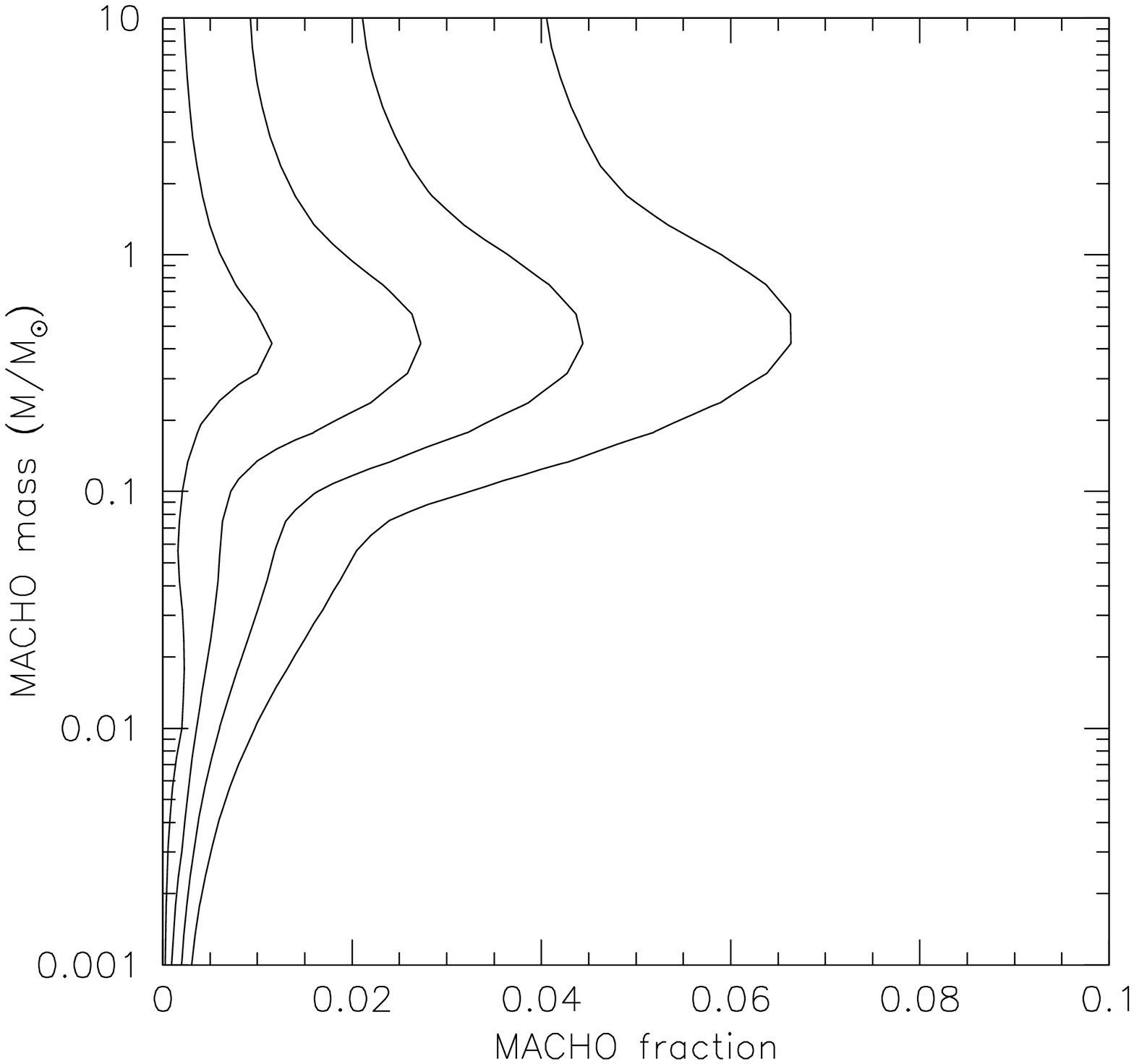, width=0.5\textwidth}
\caption{Likelihood contours for several halo fractions.  {\em Top left: 20\%,
Top right: 5\%, Bottom left: 2\%, Bottom right: 0\%}.  The contours correspond
to the usual $1\sigma$, $2\sigma$, $3\sigma$, $4\sigma$ confidence levels.  In
each case, 10 datasets were generated and the likelihoods were calculated.
Then the results were averaged.  The average results are quite representative
in each case except for the 2\% halo, where the contours for specific
realization vary significantly -- this is an indication that a 2\% halo is
about the level of sensitivity for this program.}
\label{fig:likelihood}
\end{figure*}

\begin{figure*}
\epsfig{file=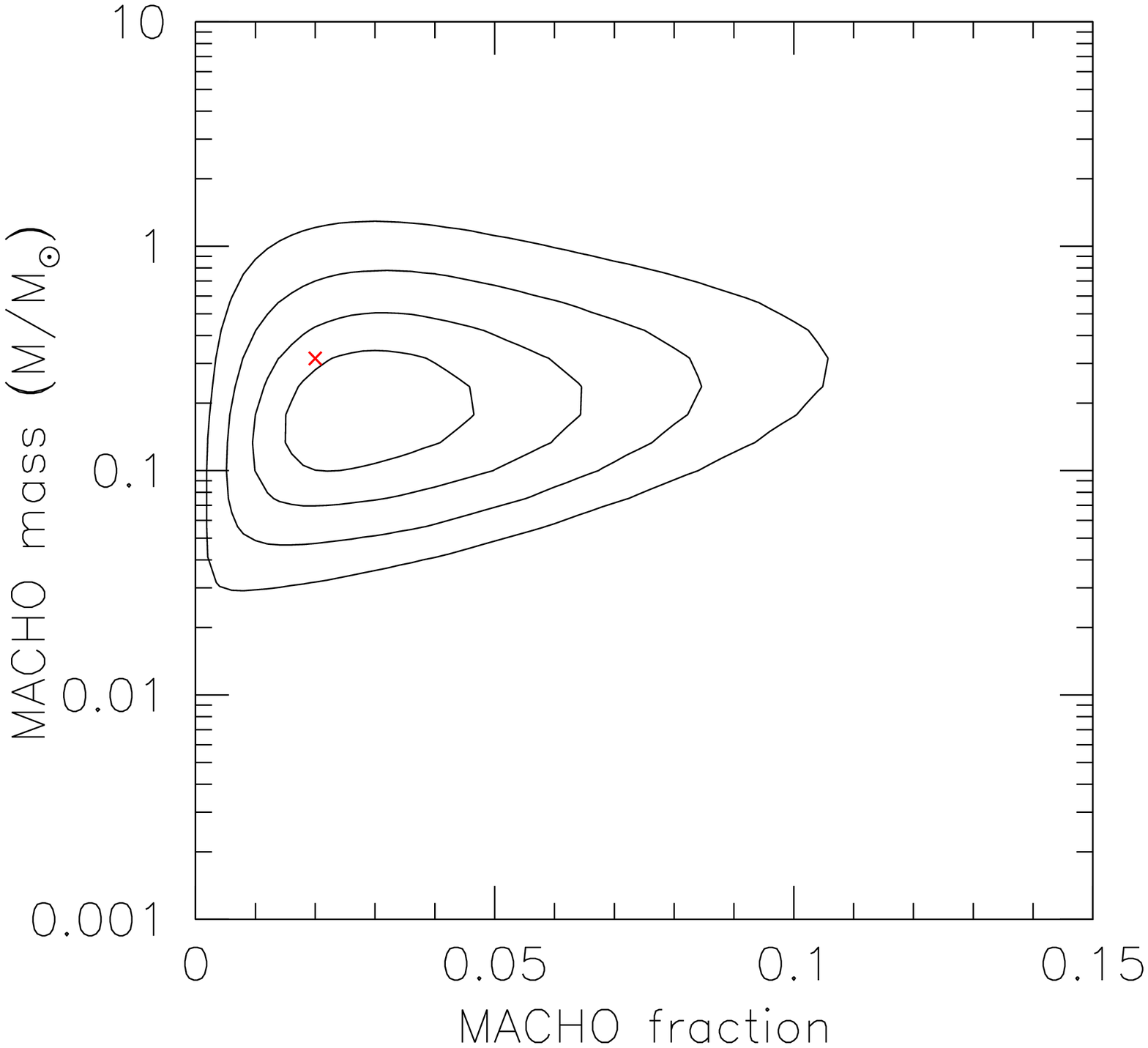, width=0.245\textwidth}
\epsfig{file=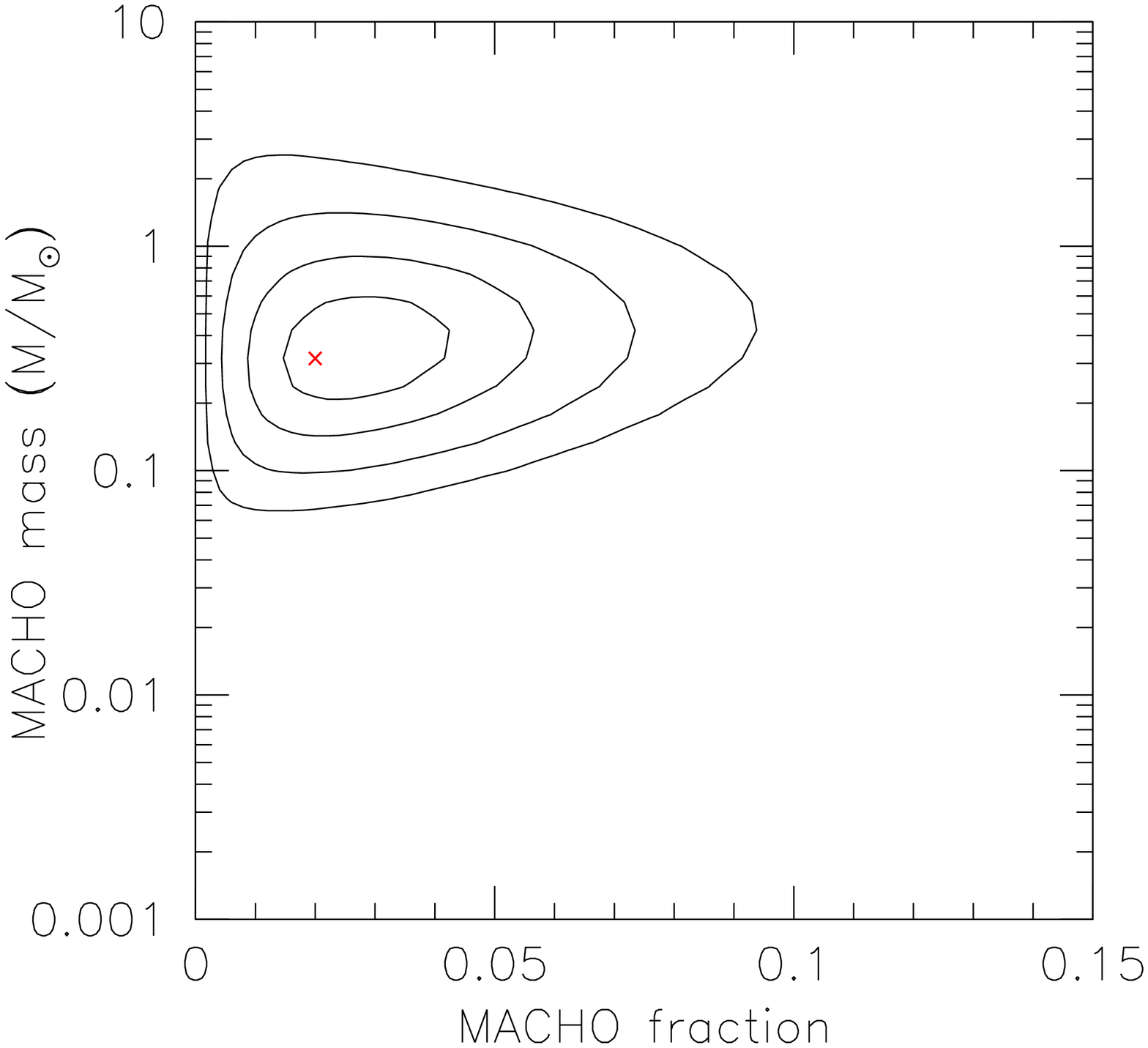, width=0.245\textwidth}
\epsfig{file=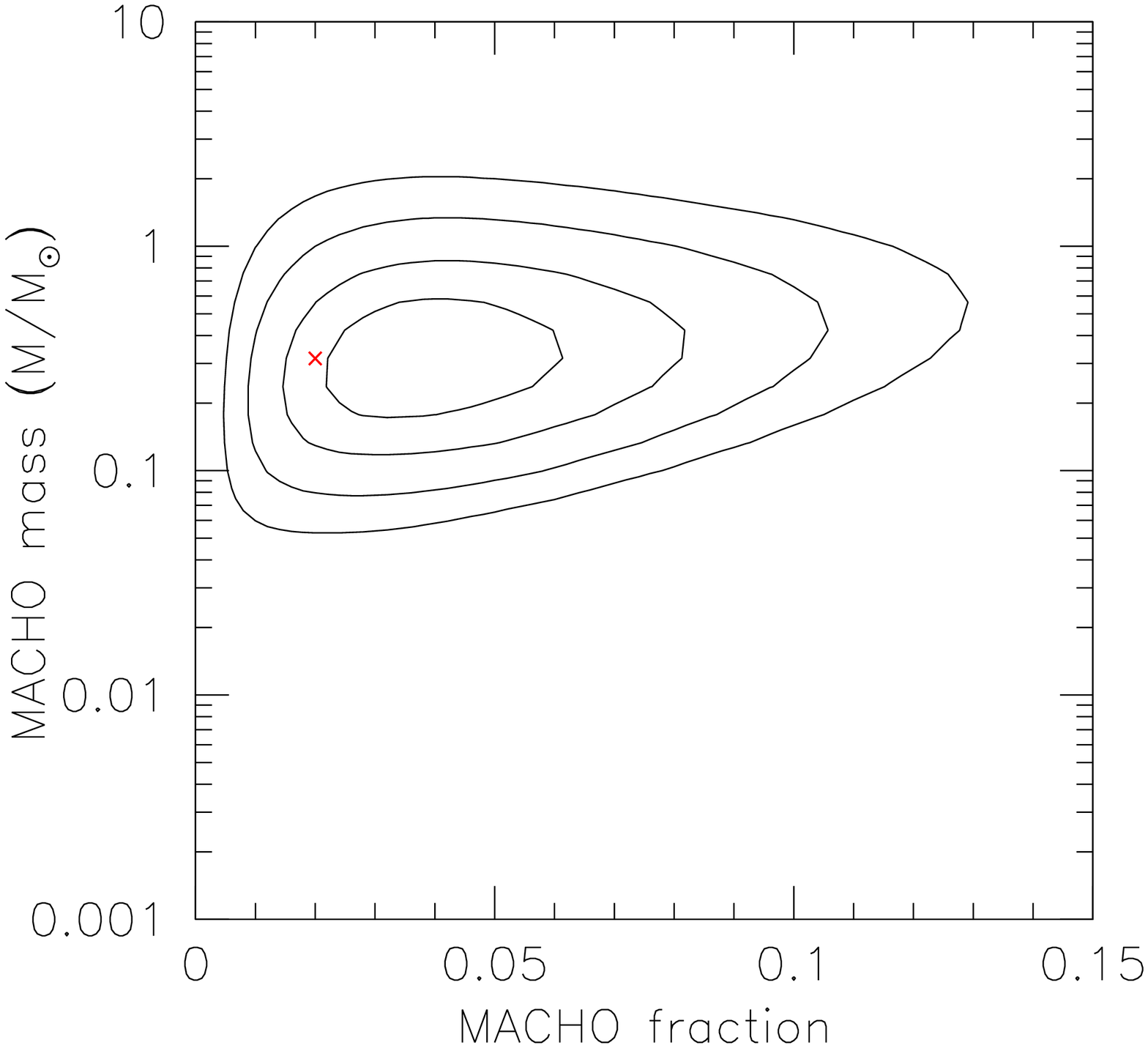, width=0.245\textwidth}
\epsfig{file=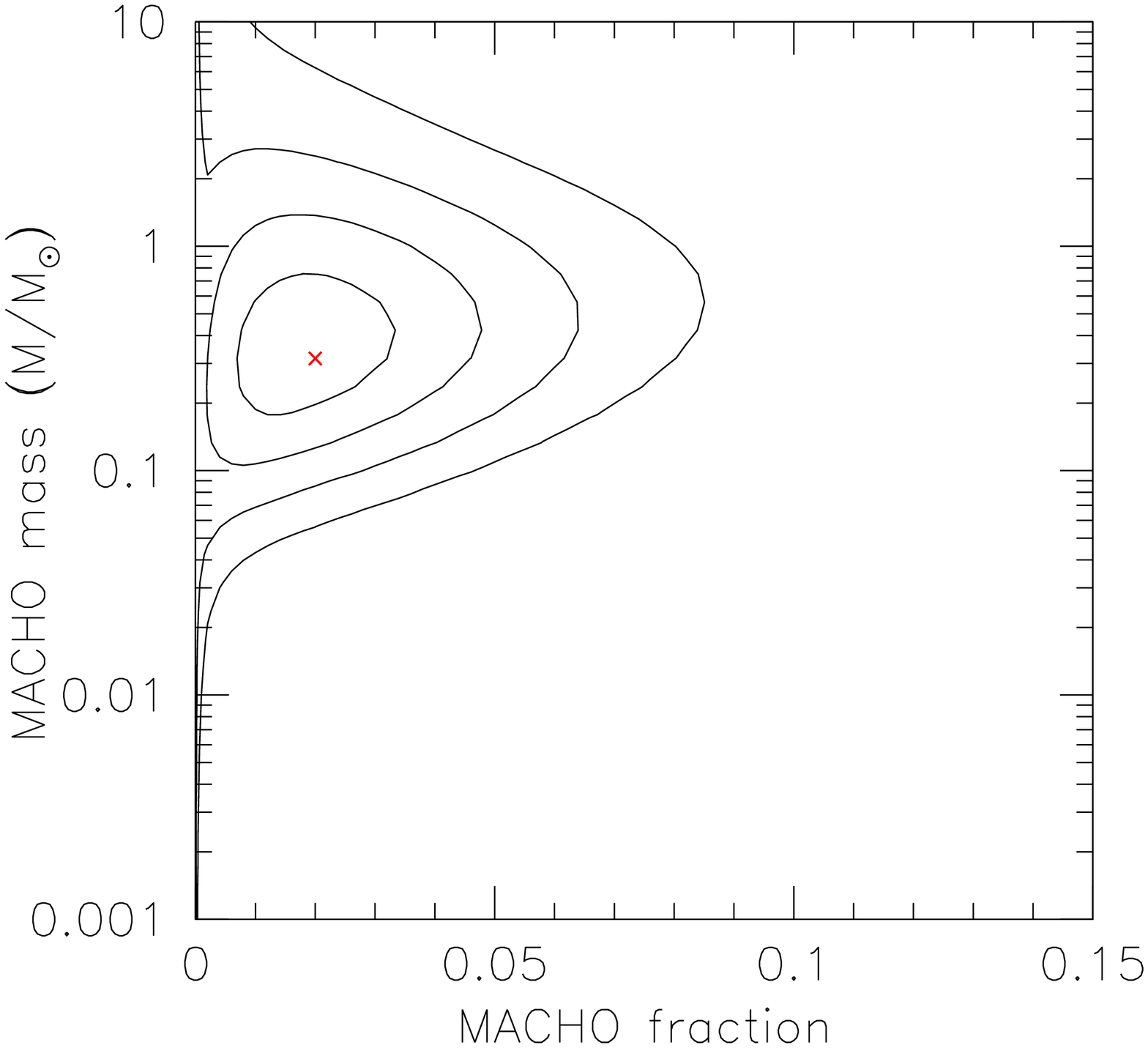, width=0.245\textwidth}
\epsfig{file=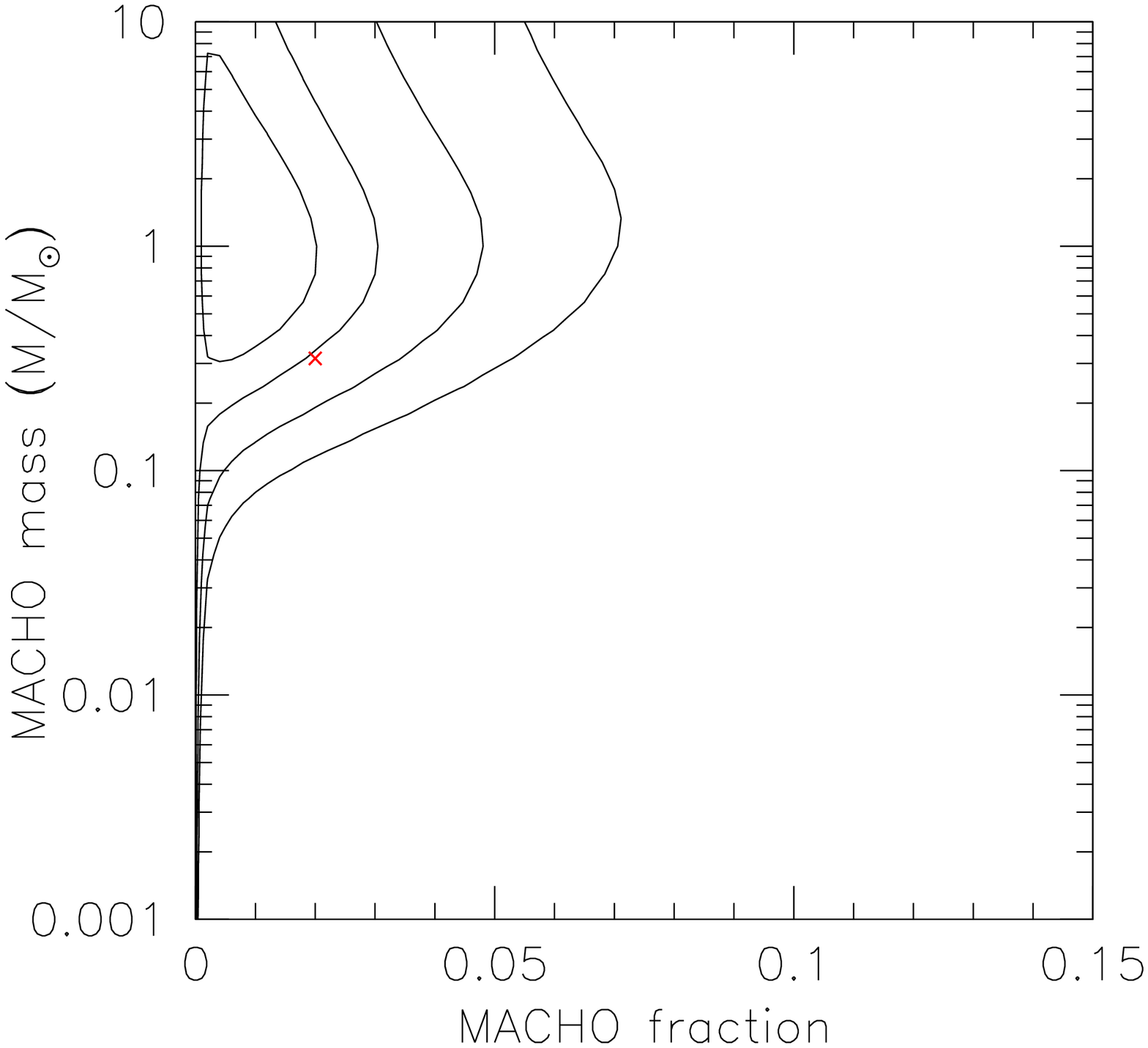, width=0.245\textwidth}
\epsfig{file=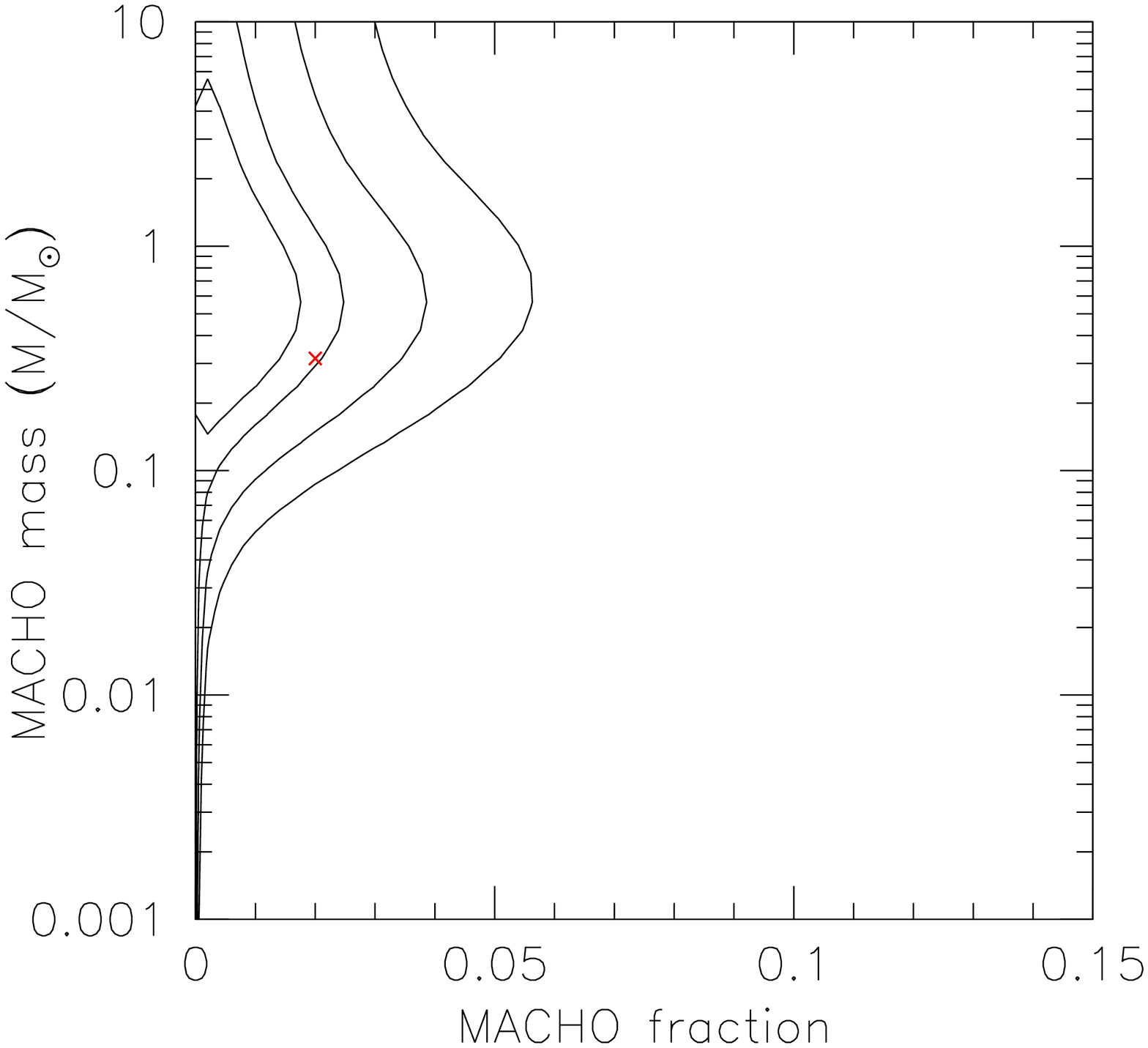, width=0.245\textwidth}
\epsfig{file=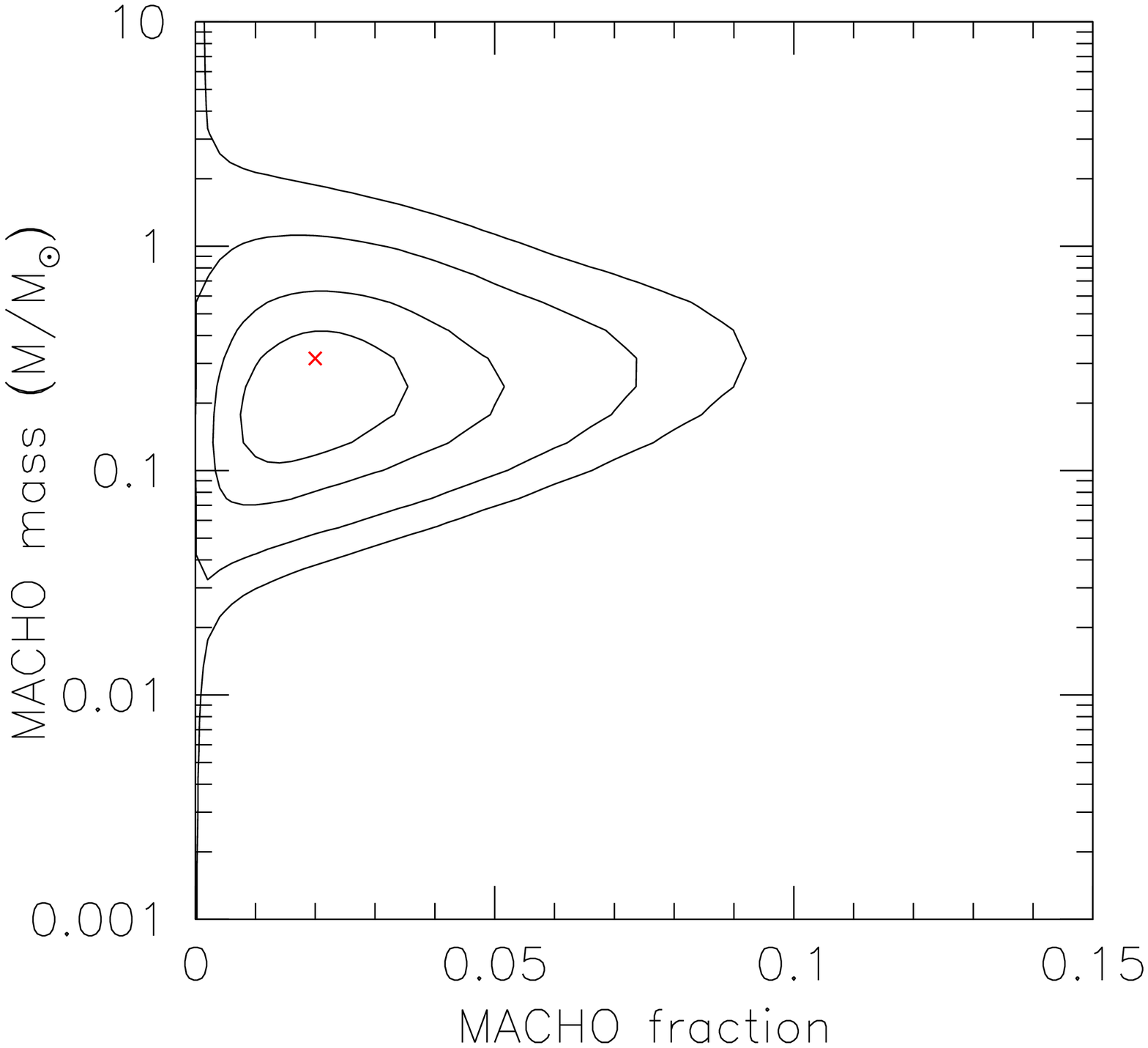, width=0.245\textwidth}
\epsfig{file=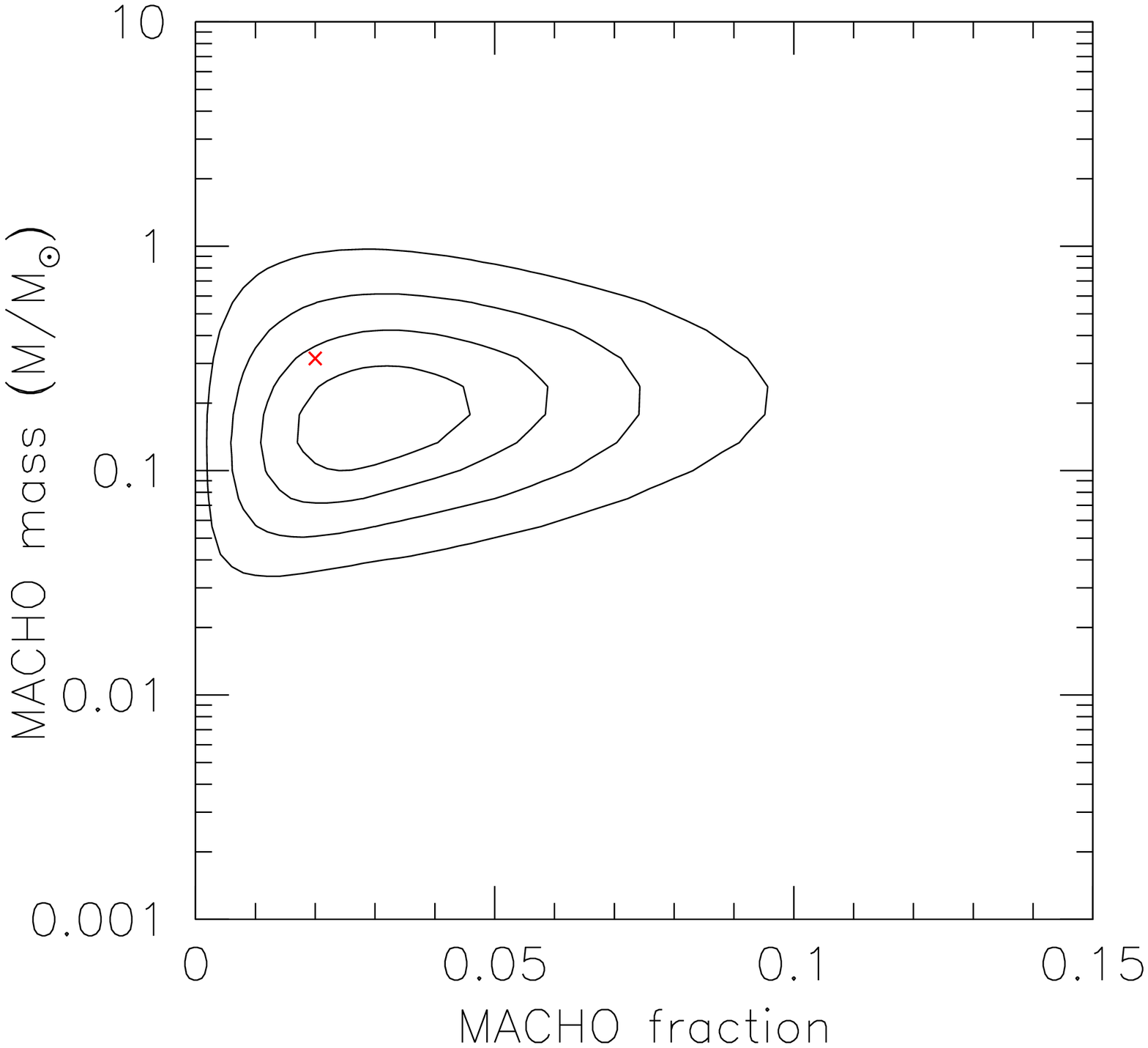, width=0.245\textwidth}
\caption{Likelihood contours for a 2\% lensing halo in several realizations.
In four cases, there is a solid detection of a lensing halo, in two others, a
weak detection, and the remaining two, no detection at all.  A solid detection
was thus found in half of the realizations.  For this reason we argue that a
2\% halo is roughly the limit that the SNAP survey could detect.}
\label{fig:lk2percent}
\end{figure*}

In Figure~\ref{fig:likelihood} we illustrate the likelihood contours for the
cases of 20\%, 5\%, 2\% and 0\% lensing halos.  For each case, 10 realizations
were generated, and likelihoods averaged with equal weight for each.  In the
case of 2\%, the likelihood contours vary significantly between realizations,
indicating that this is about the limit where a lensing halo could be
detected.  In Figure~\ref{fig:lk2percent} we illustrate several of the separate
realizations for the 2\% case.

Finally, we consider the marginalized likelihoods for lens mass and halo lens
fraction.  We can simply integrate the 2-dimensional likelihood plots along
each dimension in turn, and find the confidence intervals.  These intervals are
given in Table~\ref{tab:confidence}.

\begin{deluxetable*}{ccccccc}
\tablewidth{0pt}
\tablecaption{Confidence Intervals\label{tab:confidence}}

\tablecomments{Confidence intervals for measuring the lens mass and halo
fraction in the SNAP survey with 2 day cadence discussed in the text.  On
average, the 2\% lensing halo is not detected with high confidence, though in
several cases it was (see Figure~\ref{fig:lk2percent}).  The 20\% halo was
recovered with errors of 3.5\% in halo fraction and 0.13 dex in lens mass.  The
5\% halo was recovered with errors of 1.5\% in halo fraction and 0.20 dex in
lens mass.  In the case of no lensing halo, the 95\% limit on the halo lens
fraction was put at 2.2\%, a very considerable improvement on the capabilities
available now or in the near future.}

\tablehead{
lens fraction & lens fraction & lens fraction & lens mass &
lens mass & lens mass & lens mass \\ 
(percent) & (percent) & (percent) & (solar masses) & (solar masses) &
(dex solar) & (dex solar) \\
input & $1\sigma$ & $2\sigma$ & $1\sigma$ & $2\sigma$ & $1\sigma$ & $2\sigma$ }

\startdata
20 & 16 --- 23 & 13 --- 28 & 0.19 --- 0.35 & 0.14 --- 0.49 &
--0.72 --- --0.46 & --0.84 --- --0.31\\
5 & 3 --- 6 & 2 --- 8 & 0.19 --- 0.45 & 0.13 --- 0.73 &
 --0.73 --- --0.34 & --0.89 --- --0.14 \\
2 & 0.8 --- 3.8 & 0.0 --- 5.8 & 0.14 --- 0.51 & 0.079 --- 2.6 &
--0.85 --- --0.29 & --1.10 --- +0.42\\
0 & 0.0 --- 1.0 & 0.0 --- 2.2 & 0.017 --- 2.1 & 0.0013 --- 7.8 &
--1.76 --- +0.31 & --2.88 --- +0.89 
\enddata
\end{deluxetable*}

\section{Control of Systematics}

Wide field field microlensing surveys of M31 offer the possibility of
controlling systematic difficulties considerably more robustly than has been
possible to date.  A space based wide field survey would be especially well
suited to this task.

The relevant systematics in a microlensing survey all involve backgrounds.  The
two most important are the problems of variable star rejection and self
lensing.  Some classes of variable stars, in particular the Mira variables, can
exhibit lightcurves that mimic microlensing, and repeat only after periods of
months or years.  Identifying these classes of variable stars is thus crucial
in that false events must be avoided.  Self lensing (lensing of stars by stars)
is a significant background for microlensing surveys.  In the mock surveys
discussed here, in fact half of the events are self lensing assuming 20\%
lensing halos.  The self lensing can be separated from the halo lensing by its
spatial distribution, but this must be understood well.

General purpose multicolor wide field imagers have significant power in
understanding and accounting for these systematics.  Wide field imaging makes
for a highly homogeneous dataset.  In particular for M31, measuring the near /
far asymmetry is complicated by non-uniform time sampling, and due to
observational difficulties, the sampling of near and far side fields is rarely
even the same.  In practice this can be modeled, but this certainly introduces
a systematic error.  In addition, the presence of dust lanes spoils the
near-far symmetry of stellar backgrounds.  Multicolor observations of the
entire field, especially into the infrared, will allow this effect to be
accounted for much more robustly.  In fact, such observations, especially with
space-based resolution, will allow a more accurate self lensing model to be
constructed.

Having multiple colors, especially into the infrared, will also allow very
sophisticated identification of variable stars.  Most simply, the known classes
of troublesome variable stars tend to be red; having multiple colors for each
candidate event, including infrared colors, will allow identification of likely
variable stars with some confidence.  In particular, Miras and semiregulars
(both classes are red supergiants) are clearly identifiable in $J$-band
(accessible to a space-based survey).  Equally as important, following a
candidate microlensing event in multiple colors allows the crucial test of
gravitational lensing, that the color remains constant in time.  Variable star
outbursts usually have color variations, which would rule out the possibility
of microlensing.

A significant amount of baseline is required to identify long period variable
stars.  In a ground based survey like LSST, this will not be an issue, as at
least short exposure data will be taken for the duration of the survey, likely
several years.  A space based survey would likely need at least some ground
based support.  While it would be possible to take several exposures in each
visibility period, more complete coverage would be desirable.  Judging by event
rates, infrequent exposures of several hundred seconds on an 8 meter class
telescope would have adequate sensitivity to search for additional peaks in a
candidate microlensing lightcurve.

In the large surveys discussed in this paper, a significant number of events
are expected from stellar lenses alone.  A considerable fraction (of order
10\%) of these would be expected to display the caustic crossing signature of a
binary lens (Mao \& Paczy\'nski 1991; Baltz \& Gondolo 2001).  Caustic crossing
events exhibit unique lightcurves, thus they can be identified as microlensing
with high confidence.  The number of binary events can then be used to estimate
the expected number of single events related to the same lens population.  This
can be used as a calibration of the self lensing model.

\clearpage

\section{Discussion}

We have calculated the sensitivity to microlensing of two proposed wide field
imaging systems, and found that even with modest programs these would eclipse
anything currently available.  More importantly, such wide field multi color
surveys would address the systematic difficulties with convincingly identifying
microlensing more robustly than has been possible to date.  A space based
survey such as SNAP would be especially powerful for this purpose due to the
availability of high resolution and infrared imaging.

The chief science goal of the microlensing surveys described in this paper
would be to understand the populations of stellar--mass objects (if any)
associated with the halos of spiral galaxies.  This speaks to the issue of the
dark baryons, still unidentified, and making up roughly half of the baryon
budget in the present-day universe.

In addition to dark halo objects, other science products of such surveys would
include a census of variable stars (in several filters), and other transient
events.  One interesting possibility is the detection of binary lens events
with extreme mass ratios, indicative of giant planets.  This has been recently
achieved toward the galactic bulge with a solid detection by the OGLE and MOA
teams (Bond et al.\ 2004), followed by another detection by OGLE (Jaroszy\'nski
et al.\ 2004).  Detecting planets would necessarily involve reducing the survey
data in real time, so that microlensing alerts could be issued, and candidates
could be followed closely, either from the ground or in space.

We conclude this paper with a simple observation.  The wide field (of order
square degree) imaging systems being discussed now have obvious applications
for cosmological surveys.  In contrast, there are only a handful of objects on
the sky with angular sizes commensurate with these imagers, including the Milky
Way bulge and disk, the Magellanic Clouds, and M31.  It seems warranted that at
the very least, modest studies of these very few nearby objects should be
undertaken with the upcoming wide field technologies.

\acknowledgments

We thank Alexandre Refregier, Andrew Gould, Phil Marshall, Joseph Silk, David
Bennett, and especially Arlin Crotts for comments and suggestions, and we
gratefully acknowledge Alex Kim for providing many useful comments, and details
of the SNAP focal plane.  This work was supported in part by the
U.S. Department of Energy under contract number \mbox{DE-AC02-76SF00515}.

\end{document}